# Route to stabilized ultrabroadband microresonator-based frequency combs

Michael R.E. Lamont,[1,2,3,*] Yoshitomo Okawachi,[1] and Alexander L. Gaeta[1,3]

[1]*School of Applied and Engineering Physics, Cornell University, Ithaca, NY 14853, USA*
[2]*School of Electrical and Computer Engineering, Cornell University, Ithaca, NY 14853, USA*
[3]*Kavli Institute at Cornell for Nanoscale Science, Cornell University, Ithaca, NY 14853, USA*
*\*Corresponding author: michael.lamont@cornell.edu*

We perform the first theoretical modeling of the spectral-temporal dynamics of parametric microresonator comb generation with octave-spanning bandwidths through use of the Lugiato-Lefever model extended to include higher-order dispersion and self-steepening. We show that three distinct stages are necessary to achieve single-pulse modelocking and ultrabroadband, stabilized combs. Our simulations agree well with previous experimental demonstrations and predict many of the observed features, including multi-pulse generation, dispersive wave generation, modelocking and comb stabilization. © 2013 Optical Society of America

http://dx.doi.org/10.1364/OL.99.099999

Parametric mixing in high-$Q$ microresonators is a highly promising approach to generate optical frequency combs in a highly robust and compact platform [1-5]. Recently, such combs have been shown to exhibit modelocking and the generation of ultrashort pulses [6-8]. However, numerous aspects of the nonlinear temporal dynamics of this system are not well understood, and the modeling of these systems has been computationally intensive, which has limited numerical studies to the inclusion a few hundred modes. Herr *et al.* [8] have recently simulated mode-locking of 60 nm bandwidth (BW) frequency combs within $MgF_2$ resonators over 101 modes, showing good agreement with experiment. Experimentally, there is great interest in combs with BWs spanning an octave for applications such as optical clocks, frequency metrology, and high-precision spectroscopy. For such broadband combs in which pulse formation occurs within the microresonator, processes such as modelocking, soliton and dispersive wave generation, and the role of higher-order dispersion have not been investigated theoretically.

Since a frequency comb is composed of very narrow, discrete frequency lines spanning a broad BW, it is prohibitive to simulate all frequencies at a resolution fine enough to accurately define each comb-line. This has led several groups to apply a modal-expansion approach [8-11] in which the electric field is defined not as a function of frequency but as a summation of the modes of the resonator. This drastically reduces the computational memory necessary to simulate comb formation. However, the nonlinear Kerr term involves a triple-summation over the interacting modes, resulting in a computational time with a cubic dependence on the number of modes being considered. Thus, a frequency comb of only a few hundred modes can require extensive computing time to reach steady-state [10].

In our work, we extend the Lugiato-Lefever (LL) model [3, 9, 11, 12] to simulate the full temporal dynamics of comb generation dynamics in microresonators. This approach allows for a reduction in computational times by a factor of tens of thousands [13] and allows for simulations of modelocking of octave-spanning frequency combs. The LL model has demonstrated excellent agreement with experimental results when used to solve for a steady-state solution of the spectrum [11]. In our analysis we assume the mode frequencies are equally spaced, using the free-spectral range (FSR) adjacent to the pumped mode, and allow the phase of these modes to define the dispersion-induced frequency offset. Additionally, since we treat the field both in the time and frequency domains, additional terms such as higher-order dispersion, self-steepening, Raman scattering, two- and three-photon absorption, and free-carrier effects can easily be included in a manner similar to supercontinuum simulations [14]. Using this model, we identify distinct and necessary stages, including one that is chaotic and unstabilized, that yields a path to stable, single-pulse modelocking.

We express the LL equation with contributions from higher-order dispersion and self-steepening as,

$$T_R \frac{\partial E(t,\tau)}{\partial t} = \sqrt{\kappa}E_\text{in} + \left[-\frac{\alpha}{2} - \frac{\kappa}{2} - i\delta_0 + iL\sum_{k\geq 2}\frac{\beta_k}{k!}\left(i\frac{\partial}{\partial \tau}\right)^k + i\gamma L\left(1 + \frac{i}{\omega_0}\frac{\partial}{\partial \tau}\right)|E(t,\tau)|^2\right]E(t,\tau) \quad (1)$$

where $E$ is the resonator field, $E_\text{in}$ is the pump field at frequency $\omega_0$, $t$ and $\tau$ are the slow and fast times, respectively, describing the slow evolution of the cavity field and temporal field within the cavity at a given time $t$, $\gamma$ is the nonlinear parameter, $L$ is cavity length, $T_R$ is the round-trip time, $\delta_0$ is the cavity detuning, $\alpha$ is loss per roundtrip, $\kappa$ is the power transmission coefficient, and $\beta_k$ is the $k$-th order dispersion coefficient. Integration of Eq. (1) does not need to be in time steps equal to a roundtrip, and an adaptive step-size method [15] can be implemented.

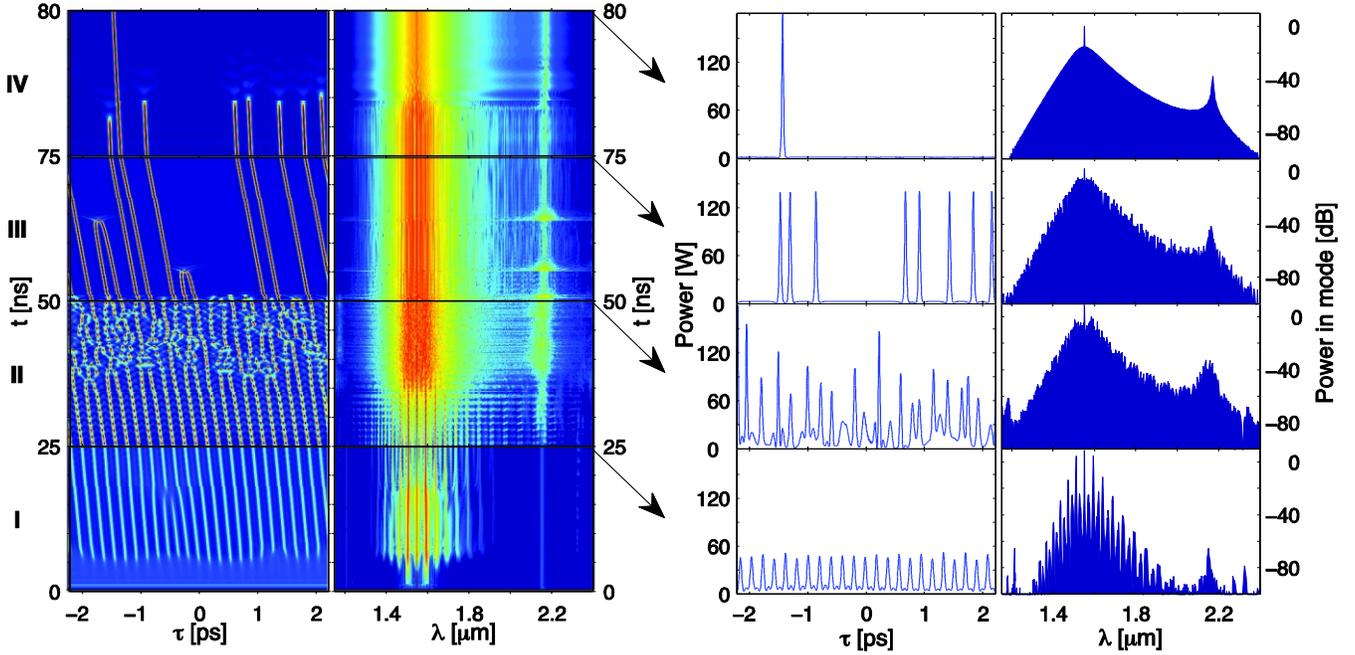

Fig. 1. Evolution of single-pulse modelocking within a 226 GHz monolithic Si$_3$N$_4$ ring resonator. Each of the four stages (bottom to top) represent an increase in pump detuning from on-resonance to $\delta_0$ = 0.02, 0.04, 0.05642 at simulated times $t$ = 25, 50, 75 ns. The pump power, $|E_{in}|^2$, is kept constant at 1.5 W. Column 1 is the temporal evolution, with power plotted on identical linear scales, and $\tau$ spanning a single roundtrip time. Column 2 is the spectral evolution, with spectral power plotted on the same 100 dB-scale. Column 3 and 4 show the temporal and spectral power, respectively, at the end of Stages I - IV.

Figure 1 shows a route to achieving stable, single-pulse modelocked octave-spanning frequency comb. The full evolution of the field consisting of 4096 modes within the microresonator is modeled starting from the initial cavity field equivalent to vacuum fluctuations is simulated for a Si$_3$N$_4$ microring resonator with a 226-GHz FSR [16] and required a total computational time of ~30 s on a notebook computer. The continuous-wave (CW) pump wavelength $\lambda_p$ at 1550 nm is positioned in the anomalous group-velocity dispersion (GVD) regime between the two zero-dispersion points (ZDP's) of the waveguide ZDP1 and ZDP2 at 1000 nm and 1760 nm, respectively, with a pump power $P_{in} = |E_{in}|^2$ of 1.5 W. The detuning of the pump field is initially set on resonance and increased at $t$ = 25, 50, 75 ns which results in the system making transitions between distinct stages of behavior that are essential for evolution to a stabilized, modelocked comb. Analogous behavior has been observed and modeled using a modal expansion approach for narrow-band combs in MgF$_2$ microresonators [8].

Stage I consists of the formation of multiple stable cavity solitons. At the bare-cavity detuning $\delta_0$ = 0, the resonance shifts away from the pump due to the Kerr effect as energy builds up within the ring. As illustrated in Fig. 2(a), anomalous GVD allows modulation instability (MI) to convert energy from the driven mode to the offset resonances ~23 modes (50 nm) away, which is dictated by phase-matching conditions [17]. Further build-up of power then causes cascaded four-wave mixing (FWM) of these MI peaks within the anomalous GVD region, seen in Fig. 2(b). Due to the low higher-order dispersion of the waveguide, ZDP2 is at a long wavelength and the possible cascaded-FWM BW is very broad. Shown in Fig. 2(c), each of these peaks also experiences MI, shifting energy to adjacent modes and creating mini-combs. During this process, FWM with the pumped mode field mirrors these mini-combs which enables the optical frequency comb to extend into the normal GVD region, as demonstrated in Fig. 2(d). Once these mini-combs begin to overlap, the temporal characteristics change dramatically, forming stable cavity solitons as initial modelocking occurs. At $t$ = 25 ns, a train of 23 pulses is seen (equal to the offset of the initial MI-driven mode) and the frequency comb shows the jagged sawtooth shape of the mini-combs. Proximity to ZDP2 results in dispersive wave formation, observable at 2150 nm. The dispersive wave arises from nearly-degenerate FWM based on a phase relationship between the soliton and the dispersive radiation [18].

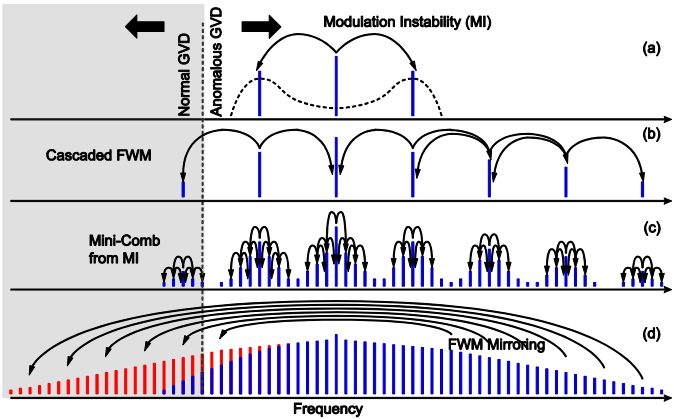

Fig. 2. Processes involved in initial comb formation. (a) Anomalous GVD allows for modulation instability (MI) from pump. (b) Cascaded FWM from further power build-up. (c) Mini-comb formation due to MI from each of the cascaded peaks. (d) FWM mirroring with central pumped mode.

In Stage II, as the pump is tuned closer to the Kerr-shifted resonance, the intracavity power increases, and the multi-pulse modelocking is lost due to instability of the cavity solitons. The pulses exhibit periodic amplitude fluctuations, similar to higher-order solitons, causing periodic broad spectral features. Adjacent pulses interact with each other causing destabilization and what appears to be chaotic behavior [19]. At $t = 50$ ns, the frequency comb is at its broadest, including increased power in the dispersive wave, however the time-domain is now predominantly a series of random pulses.

Further detuning of the pump past the Kerr-shifted resonance in Stage III causes the intracavity power to drop, and modelocking is again observed. After the change in $\delta_0$, multiple cavity solitons re-form, but only the stable ones persist. The remaining pulses collide with the nearest neighbors and annihilate, causing spikes in power and broad lines seen in the corresponding spectral evolution. At $t = 75$ ns, six pulses remain with apparent random spacing. The frequency comb has decreased slightly in both total power and BW.

In the final stage, the pump detuning is increased again. At the lower intracavity powers the remaining cavity solitons destabilize and dissipate. At the suitable pump detuning, the system evolves to a single pulse within the microresonator with a full-width at half-maximum pulsewidth of 32 fs. The modelocked frequency comb is smooth and spans an octave across ZDP2 into normal GVD regime, as has been observed experimentally [16], with the dispersive wave feeding energy into the longer wavelength modes. Very fine tuning of this final pump detuning is necessary for a single pulse to survive, since only the most stable cavity soliton remains. However, once single-pulsing is achieved, the pump detuning can be reduced to increase stability. Again, this is consistent with experimental observation in which single-pulse modelocking has been observed [6-8].

Figure 4(a) shows the spectrogram of the single-pulse modelocked frequency comb on a logarithmic scale. We observe the residual pump coupled into the resonator as a CW pedestal around the pulse near 1550 nm. If $\delta_0$ is reduced after Stage IV to increase the stability of the system, the CW pedestal also increases. A dispersive wave component near 2150 nm is also evident. Although the modes at longer wavelengths contribute to the pulse, the pulse associated with the dispersive wave experiences temporal walk-off within the ring. This suggests that the dispersive wave pulse near 2150 nm is not modelocked. This may be an issue when these modes are frequency-doubled for $f-2f$ self-referencing for frequency stabilization. The theoretical spectrum of the final modelocked frequency comb is in good agreement with the experimental spectrum in Fig. 4 (b). This single-pulse frequency comb spectrum also matches the steady-state solution of the same system presented by Coen et al. [11].

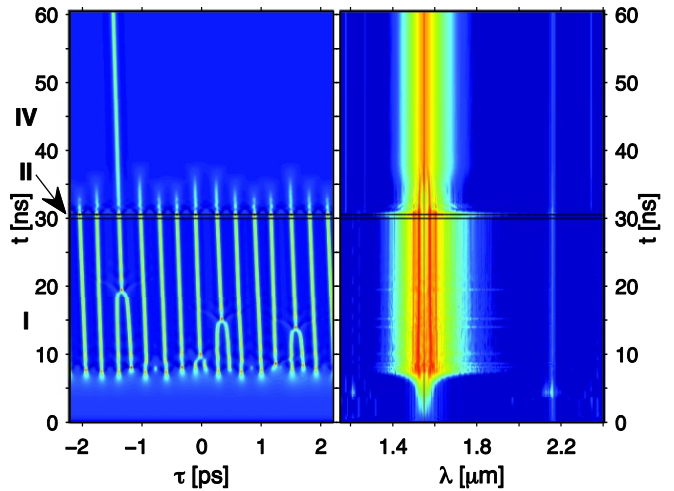

Fig. 3. Temporal and spectral evolution with the same parameters used in Fig. 1. For this case $\delta_0$ is kept at 0.015 and the pump power is set at 0.5 W for 30 ns, then increased to 1 W for 0.5 ns, and dropped to 0.376 W for the last 30 ns, demonstrating a different route to single-pulse modelocking.

There are a few keys points to note concerning the route to single-pulse modelocking. Progressing through Stage II is a necessary condition for reaching this regime, and it can only occur if the supplied pump power is high enough to support fundamental soliton conditions for the given GVD. Stage III differs greatly from Stage I, although both support stable cavity solitons. In the first stage, a reduction of power does not result in individual pulses vanishing. Instead, the magnitude (peak-to-valley) of all the pulses is reduced until oscillation in all modes is lost. While in Stage II or III, a reduction in pump power progresses the system toward single-pulse modelocking. Figure 3 highlights the necessity of the system passing through Stage II to achieve single-pulse modelocking. Using the same parameters as those used in Fig. 1, the pump detuning $\delta_0$ is kept constant at 0.015 and only $P_{in}$ is varied. Starting with $P_{in} = 0.5$ W, multiple cavity solitons form in the same manner as Stage I of Fig. 1. In Stage II, the power is increased to 1 W for only 0.5 ns before decreasing to 0.376 W, below the power required for initial oscillation. With less power, all but a single cavity soliton dissipates similar to Stage IV in Fig. 1 and modelocking is achieved. If the system does not proceed to Stage II, for example, $P_{in}$ is simply decreased from 0.5 W to 0.376 W, then oscillation is lost as previously described. Our studies indicate that Stage II cannot be reached if $\delta_0 = 0$.

Lastly, we consider the effects of dispersion on the comb BW, shown in Fig. 5. Recently, Coen et al. performed analysis which was able to predict the comb BW for a number of cases [20]. Similarly, Herr *et al.* estimate the minimal soliton width [8]. In Fig. 5, we show a case, as noted by the authors, for which the prediction deviates for lower $\beta_2$ when higher-order dispersion becomes important. Each simulation is run under the same conditions as Fig. 1, with the core width of the waveguide varying from 1800 nm to 2600 nm, which shifts the ZDP's and the amount of anomalous GVD at the pump wavelength. Decreasing the anomalous GVD at the pump [Fig. 5(a) to (c)], causes the offset of the initial oscillating mode [Fig. 2(a)] to increase and the frequency comb BW to broaden. Note that the frequency comb in Fig. 5(c) extends into the normal GVD regions at both high and low frequencies, with the modes near ZDP1 mirroring modes across the ZDP as shown Fig. 2(d). The BW of the comb decreases from Fig. 2(c) to (d), despite the decrease in anomalous GVD due to the proximity of ZDP1 (at higher frequency) hindering the FWM cascade. Although the GVD remains anomalous at $\lambda$ = 1550 nm for waveguide widths up to 2750 nm, single-pulse modelocking was not possible with $P_{in}$ = 1.5 W for waveguides with widths larger than 2600 nm. In these cases, the fundamental soliton condition could not be met. These results suggest that the optimal anomalous GVD for a broadband optical frequency combs would be low and flat, that is, should have small negative $\beta_2$ and small higher-order dispersion such that ZDP1 and ZDP2 are widely separated. However, the characteristic shape of the waveguide GVD contribution limits what is possible by dispersion-engineering step-index waveguides [21]. We note that as ZDP2 shifts close to the pump mode [Fig. 5(c) and (d)], the center of the frequency comb moves to higher frequency modes. This may be due to the relative efficiencies of the FWM processes shown in Fig. 2(b) and (d) shifting the comb away from normal GVD regions.

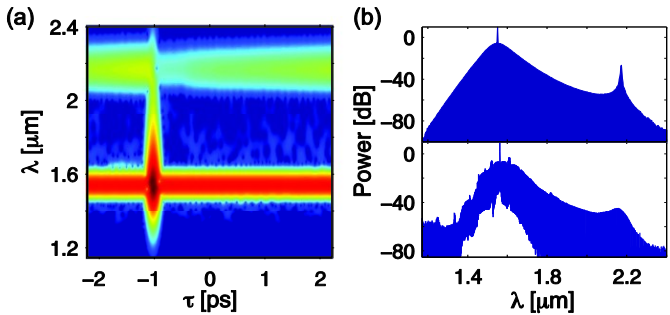

Fig. 4. (a) Spectrogram of the single-pulse modelocked frequency comb, plotted on a 100 dB-scale. (b) Comparison of the theoretical (top) and experimental (bottom) frequency comb spectra. Data is from Okawachi et al. [16].

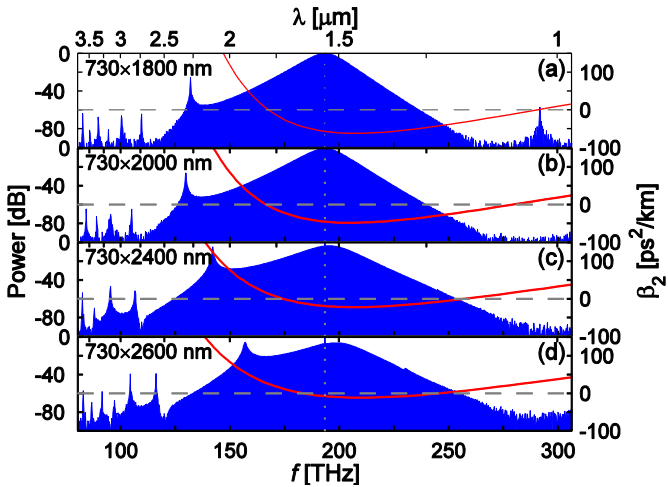

Fig. 5. Frequency comb spectra (blue area) of waveguide with varying widths pumped at $\lambda_p$ = 1550 nm. All other materials and dimensions are the same. The group-velocity dispersion for each waveguide geometry is shown by the red (solid) line.

In conclusion, we perform extensive modeling of the temporal dynamics of parametric comb formation and pulse modelocking within a microresonator with comb BWs spanning an octave. We find that achieving a stable, modelocked comb requires a multi-stage detuning process of the pump wave in which the system must pass through a key regime of aperiodic multi-pulse behavior. In the last stage, fine control of detuning or pump power is necessary achieve single-pulse modelocking, which is similar to that observed with narrowband combs generated in $MgF_2$ microresonators [8]. The predicted spectra compare well with previous experimental measurements and agree with steady-state solutions [11]. We also show that small, flat anomalous GVD can result in broader BW frequency combs. This model for the temporal dynamics appears to agree well with those based on a modal expansion and allows for a few orders of magnitude reduction in computational time and allows for thousands of modes to be included, a necessity for broadband frequency combs.

This work was supported by DARPA via the QuASAR program and AFOSR under grant FA9550-12-1-0377.

### References

1. T. J. Kippenberg, R. Holzwarth, and S. A. Diddams, Science **332,** 555-559 (2011).
2. P. Del'Haye, A. Schliesser, O. Arcizet, T. Wilken, R. Holzwarth, and T. J. Kippenberg, Nature **450,** 1214-1217 (2007).
3. A. B. Matsko, A. A. Savchenkov, W. Liang, V. S. Ilchenko, D. Seidel, and L. Maleki, Opt. Lett. **36,** 2845-2847 (2011).
4. F. Ferdous, H. Miao, D. E. Leaird, K. Srinivasan, J. Wang, L. Chen, L. T. Varghese, and A. M. Weiner, Nat. Photonics **5,** 770-776 (2011).
5. S. B. Papp and S. A. Diddams, Phys. Rev. A **84,** 053833 (2011).
6. M. A. Foster, J. S. Levy, O. Kuzucu, K. Saha, M. Lipson, and A. L. Gaeta, arXiv:1102.0326.
7. K. Saha, Y. Okawachi, B. Shim, J. S. Levy, R. Salem, A. R. Johnson, M. A. Foster, M. R. Lamont, M. Lipson, and A. L. Gaeta, Opt. Express **21,** 1335-1343 (2013).
8. T. Herr, V. Brasch, J. D. Jost, C. Y. Wang, N. M. Kondratiev, M. L. Gorodetsky, and T. J. Kippenberg, arXiv:1211.0733 [physics.optics] (2013).
9. M. Haelterman, S. Trillo, and S. Wabnitz, Opt. Commun. **91,** 401-407 (1992).
10. Y. K. Chembo and N. Yu, Phys. Rev. A **82,** 033801 (2010).
11. S. Coen, H. G. Randle, T. Sylvestre, and M. Erkintalo, Opt. Lett. **38,** 37-39 (2013).
12. L. A. Lugiato and R. Lefever, Phys. Rev. Lett. **58,** 2209-2211 (1987).
13. Y. K. Chembo and C. R. Menyuk, arXiv:1210.8210 [physics.optics] (2012).
14. J. M. Dudley, G. Genty, and S. Coen, Rev. Mod. Phys. **78,** 1135-1184 (2006).
15. O. V. Sinkin, R. Holzlohner, J. Zweck, and C. R. Menyuk, J. Lightwave Technol. **21,** 61-68 (2003).
16. Y. Okawachi, K. Saha, J. S. Levy, Y. H. Wen, M. Lipson, and A. L. Gaeta, Opt. Lett. **36,** 3398-3400 (2011).
17. T. Herr, K. Hartinger, J. Riemensberger, C. Y. Wang, E. Gavartin, R. Holzwarth, M. L. Gorodetsky, and T. J. Kippenberg, Nat. Photonics **6,** 480-487 (2012).
18. M. A. Foster, A. C. Turner, M. Lipson, and A. L. Gaeta, Opt. Express **16,** 1300-1320 (2008).
19. A. B. Matsko, W. Liang, A. A. Savchenkov, and L. Maleki, Opt.Lett. **38,** 525-527 (2013).
20. S. Coen and M. Erkintalo, Opt. Lett. **38,** 1790 (2013).
21. M. R. E. Lamont, B. T. Kuhlmey, and C. M. de Sterke, Opt. Express **16,** 7551-7563 (2008).

# References

1. T. J. Kippenberg, R. Holzwarth, and S. A. Diddams, "Microresonator-based optical frequency combs," Science **332,** 555-559 (2011).
2. P. Del'Haye, A. Schliesser, O. Arcizet, T. Wilken, R. Holzwarth, and T. J. Kippenberg, "Optical frequency comb generation from a monolithic microresonator," Nature **450,** 1214-1217 (2007).
3. A. B. Matsko, A. A. Savchenkov, W. Liang, V. S. Ilchenko, D. Seidel, and L. Maleki, "Mode-locked Kerr frequency combs," Opt. Lett. **36,** 2845-2847 (2011).
4. F. Ferdous, H. Miao, D. E. Leaird, K. Srinivasan, J. Wang, L. Chen, L. T. Varghese, and A. M. Weiner, "Spectral line-by-line pulse shaping of on-chip microresonator frequency combs," Nat. Photonics **5,** 770-776 (2011).
5. S. B. Papp and S. A. Diddams, "Spectral and temporal characterization of a fused-quartz-microresonator optical frequency comb," Phys. Rev. A **84,** 053833 (2011).
6. M. A. Foster, J. S. Levy, O. Kuzucu, K. Saha, M. Lipson, and A. L. Gaeta, "A silicon-based monolithic optical frequency comb source," arXiv:1102.0326.
7. K. Saha, Y. Okawachi, B. Shim, J. S. Levy, R. Salem, A. R. Johnson, M. A. Foster, M. R. Lamont, M. Lipson, and A. L. Gaeta, "Modelocking and femtosecond pulse generation in chip-based frequency combs," Opt. Express **21,** 1335-1343 (2013).
8. T. Herr, V. Brasch, J. D. Jost, C. Y. Wang, N. M. Kondratiev, M. L. Gorodetsky, and T. J. Kippenberg, "Mode-locking in an optical microresonator via soliton formation," arXiv:1211.0733 [physics.optics] (2013).
9. M. Haelterman, S. Trillo, and S. Wabnitz, "Dissipative Modulation Instability in a Nonlinear Dispersive Ring Cavity," Opt. Commun. **91,** 401-407 (1992).
10. Y. K. Chembo and N. Yu, "Modal expansion approach to optical-frequency-comb generation with monolithic whispering-gallery-mode resonators," Phys. Rev. A **82,** 033801 (2010).
11. S. Coen, H. G. Randle, T. Sylvestre, and M. Erkintalo, "Modeling of octave-spanning Kerr frequency combs using a generalized mean-field Lugiato-Lefever model," Opt. Lett. **38,** 37-39 (2013).
12. L. A. Lugiato and R. Lefever, "Spatial Dissipative Structures in Passive Optical-Systems," Phys. Rev. Lett. **58,** 2209-2211 (1987).
13. Y. K. Chembo and C. R. Menyuk, "Spatiotemporal Lugiato-Lefever formalism for Kerr comb generation in whispering-gallery-mode resonators," arXiv:1210.8210 [physics.optics] (2012).
14. J. M. Dudley, G. Genty, and S. Coen, "Supercontinuum generation in photonic crystal fiber," Rev. Mod. Phys. **78,** 1135-1184 (2006).
15. O. V. Sinkin, R. Holzlohner, J. Zweck, and C. R. Menyuk, "Optimization of the split-step Fourier method in modeling optical-fiber communications systems," J. Lightwave Technol. **21,** 61-68 (2003).
16. Y. Okawachi, K. Saha, J. S. Levy, Y. H. Wen, M. Lipson, and A. L. Gaeta, "Octave-spanning frequency comb generation in a silicon nitride chip," Opt. Lett. **36,** 3398-3400 (2011).
17. T. Herr, K. Hartinger, J. Riemensberger, C. Y. Wang, E. Gavartin, R. Holzwarth, M. L. Gorodetsky, and T. J. Kippenberg, "Universal formation dynamics and noise of Kerr-frequency combs in microresonators," Nat. Photonics **6,** 480-487 (2012).
18. M. A. Foster, A. C. Turner, M. Lipson, and A. L. Gaeta, "Nonlinear optics in photonic nanowires," Opt. Express **16,** 1300-1320 (2008).
19. A. B. Matsko, W. Liang, A. A. Savchenkov, and L. Maleki, "Chaotic dynamics of frequency combs generated with continuously pumped nonlinear microresonators," Opt. Lett. **38,** 525-527 (2013).
20. S. Coen and M. Erkintalo, "Universal scaling laws of Kerr frequency combs," Opt. Lett. **38,** 1790 (2013).
21. M. R. E. Lamont, B. T. Kuhlmey, and C. M. de Sterke, "Multi-order dispersion engineering for optimal four-wave mixing," Opt. Express **16,** 7551-7563 (2008).